\documentclass[conference]{IEEEtran}
\ifCLASSINFOpdf
\else
\fi
\usepackage[
  margin=2in,
  pass,% keep layout unchanged 
]{geometry}

\usepackage{amssymb} %,amsmath}
\usepackage[usenames,dvipsnames,svgnames,x11names,table]{xcolor} % http://en.wikibooks.org/wiki/LaTeX/Colors

\definecolor{VeryLightGray}{rgb}{0.92,0.92,0.92}

\usepackage[framemethod=tikz]{mdframed}
\newmdenv[
  innerleftmargin=7pt,
  innerrightmargin=7pt,
  tikzsetting={draw=black,dashed,line width=0.5pt,dash pattern = on 4pt off 2pt},
  linecolor=white,
  backgroundcolor=white
]{dashedbox}
\newmdenv[
  innerleftmargin=7pt,
  innerrightmargin=7pt,
  tikzsetting={draw=black, line width=0.5pt},
  linecolor=black,
  backgroundcolor=white
]{normalbox}

\usepackage{multirow} 
\usepackage{bm} 
\usepackage{enumitem}
\usepackage{calc}
\usepackage{subcaption}

\begin{document}
%\title{LivelySketches: Supporting the Analog and Digital Lifecycle of Sketches in Software Development}
%\title{Round-Trip Sketches: Supporting Sketch Lifecycles from Analog to Digital and Back}
\title{Round-Trip Sketches: Supporting the Lifecycle of Software Development Sketches from\\ Analog to Digital and Back}

\author{\IEEEauthorblockN{Sebastian Baltes, Fabrice Hollerich, and Stephan Diehl}
\IEEEauthorblockA{Department of Computer Science\\
University of Trier\\
Trier, Germany\\
Email: research@sbaltes.com, diehl@uni-trier.de}}


%Telephone: (800) 555--1212, Fax: (888) 555--1212}


\maketitle

\begin{abstract}
Sketching is an important activity for understanding, designing, and communicating different aspects of software systems such as their requirements or architecture. 
Often, sketches start on paper or whiteboards, are revised, and may evolve into a digital version.
Users may then print a revised sketch, change it on paper, and digitize it again.
Existing tools focus on a paperless workflow, i.e., archiving analog documents, or rely on special hardware---they do not focus on integrating digital versions into the analog-focused workflow that many users follow.
In this paper, we present the conceptual design and a prototype of \textit{LivelySketches}, a tool that supports the ``round-trip'' lifecycle of sketches from analog to digital and back.
The proposed workflow includes capturing both analog and digital sketches as well as relevant context information. 
In addition, users can link sketches to other related sketches or documents.
They may access the linked artifacts and captured information using digital as well as augmented analog versions of the sketches.
We further present results from a formative user study with four students and outline possible directions for future work.
\end{abstract}


% \fi
\IEEEpeerreviewmaketitle

\section{Introduction}

Sketches and diagrams play an important role in design-related activities~\cite{Sonnentag98, Tversky03, Black90}.
Artists sketch to clarify existing ideas and to develop new ones~\cite{Fish90}.
In mechanical design, sketches not only document final designs, but also provide designers with a memory extension to help ideas taking shape and to communicate concepts to colleagues~\cite{Ullman90}.
Beside sketches being an external representation of memory and a means for communication~\cite{Tversky02, Tversky01}, they serve as documentation~\cite{Schuetze03}.
The ambiguity in sketches is a source of creativity~\cite{Goldschmidt03} and they support problem solving and understanding~\cite{Suwa02}.
In engineering, controlled experiments have shown that the possibility to sketch has a positive effect on the quality of the solution~\cite{Schuetze03}.
Software developers use sketches and diagrams to understand, to design, and to communicate different aspects of software systems~\cite{Dekel07, Cherubini07, Walny11, Baltes14}.
Most software engineering sketches do not follow formal conventions like the \emph{Unified Modeling Language} (UML), but have an informal, ad-hoc nature~\cite{Cherubini07, Petre13, Dekel07, Baltes14, Gorschek14}.

Media used for sketch creation include not only whiteboards and paper, but also software tools like Photoshop and PowerPoint~\cite{Walny11, Cherubini07, Myers08, Gorschek14, Baltes14}.
Often, sketches are revised~\cite{Baltes14} and pass through transitions from analog to digital media~\cite{Walny11}, because digital sketches can more easily be edited, copied, organized, and shared~\cite{Willett15}.
Even if a digital version exists, analog sketches may be kept as a memory aid~\cite{Walny11a}.
Context information is often needed to understand informal sketches~\cite{LaToza06} and information may get lost due to the transient nature of sketches~\cite{Cherubini07, Baltes14}.

Despite the widespread usage of sketches in many domains, to the best of our knowledge there is currently no tool that explicitly supports the complete analog and digital lifecycle of sketches.
Popular tools like \emph{EverNote} and \emph{OneNote} focus on a paperless workflow, i.e., archiving analog documents, not on integrating digital versions into the analog-focused workflow that many users follow~\cite{Baltes14, Walny11a}.
Other proposed tools rely on special hardware like digital pens and compatible paper for creating analog sketches~\cite{Guimbretiere03, Liao06, Liao08, Weibel08}, need a special scanning device to access digital versions~\cite{Norrie03}, or treat analog documents only as passive link anchors for digital resources~\cite{Bian11}.
%Because such special hardware is not available in many situations, our approach only relies on tablets and sticky labels with QR codes, no special paper or pen is needed (see Sections \ref{sec:conceptual-design} and \ref{sec:prototype-implementation}). Moreover, we treat analog documents as first-order artifact and not only as passive link anchors for digital resources..

From the literature cited above and in particular our own research on the use of sketches and diagrams in software engineering practice~\cite{Baltes14}, we derived four main requirements that a tool supporting the analog and digital lifecycle of sketches should implement: (1)~it should be possible to archive analog sketches along with context information needed for their understanding, (2)~the tool should provide a version control for evolving sketches, (3)~since sketches are usually connected and embedded in a work context, it should be possible to link them to other sketches or related artifacts, and (4)~the tool should support the ``round-trip'' of sketches from analog to digital and back.
In the following, we further elaborate on the conceptual design, present a prototype implementation of this concept named \textit{LivelySketches}, and report findings from a formative user study.

\section{Conceptual Design}
\label{sec:conceptual-design}

\begin{figure*}[!t]
\centering
\includegraphics[width=0.98\textwidth]{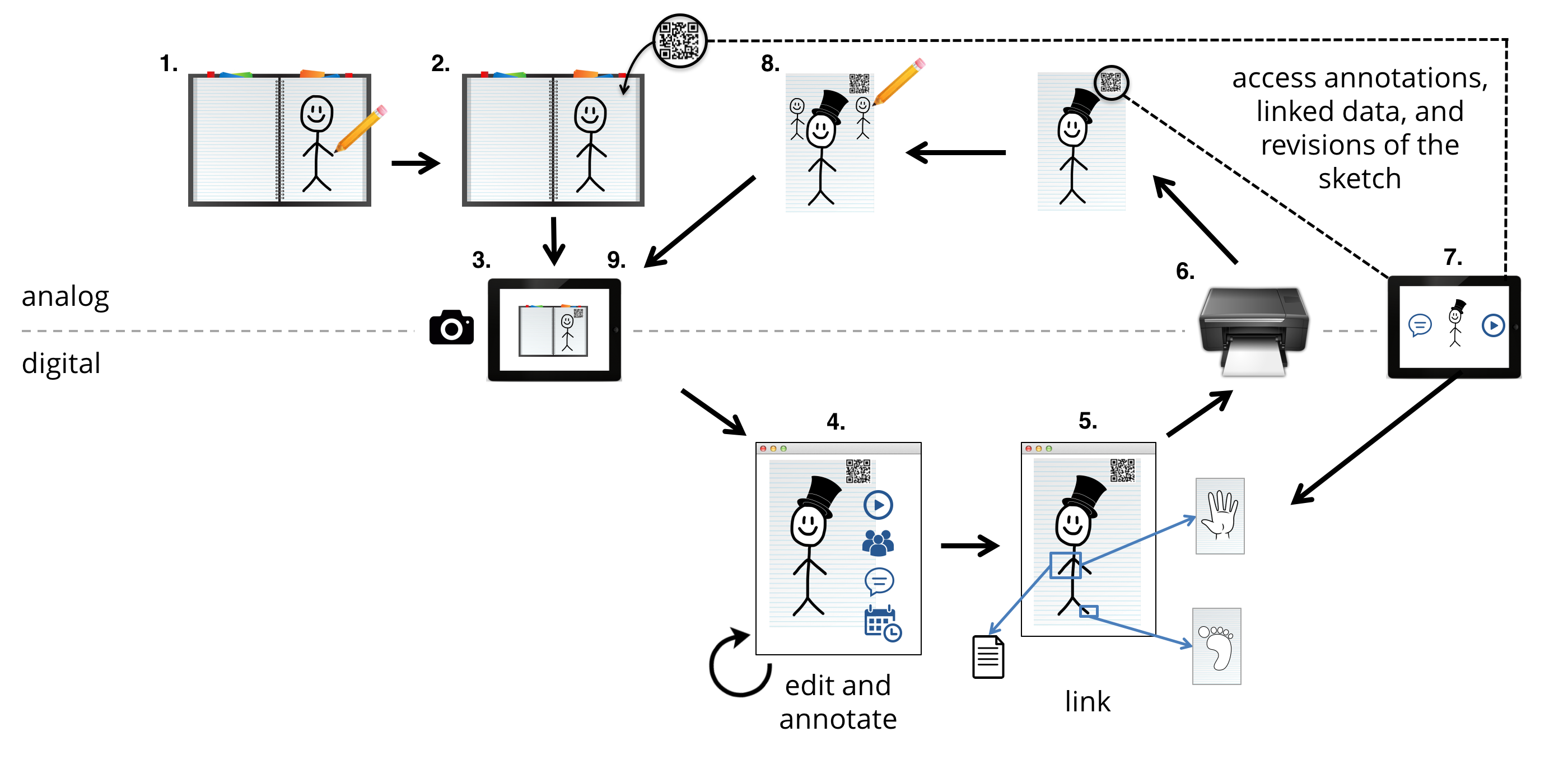}
\caption{The conceptual workflow of round-trip sketching from analog to digital and back.}
\label{fig:lifecycle}
\end{figure*}

Sketches often start on analog media like paper or whiteboards and later get digitized to share and revise them~\cite{Baltes14, Cherubini07, Walny11a}.
However, sketches do not only evolve digitally, but may be printed out or redrawn on paper or whiteboards.
We denote this process involving transitions from analog to digital sketching and vice versa as ``round-trip sketching''.
In this section, we describe an exemplary scenario and a conceptual workflow to derive requirements for a tool providing support for round-trip sketching.

In our scenario, a software development team discusses possible extensions of an app.
They collect emerging ideas in a shared whiteboard sketch.
Furthermore, some developers write down their own thoughts in personal notebooks.
At some point, the whiteboard becomes too cluttered and the team leader decides to clean the whiteboard partially.
After the meeting, one developer is asked to create a polished version of the whiteboard sketch for the customer.
This sketch is also posted in an internal development wiki.
A week later, the team uses a printout of the polished sketch together with the developers' notes to continue where they left off.

In this scenario, several issues arise:
When the whiteboard is cleaned, it is not possible to go back to erased content that may still be relevant to understand the evolution of the sketch.
Further, the developers' notes are related to the shared sketch, but the connection is only present in their mind.
%Moreover, sketches are not isolated but may be related to other sketches or documents.
To be able to refine the whiteboard sketch, the developer may need contextual information that has not been captured during the meeting.
If the sketch is shared with others (e.g., in a wiki), parallel versions of the sketch may evolve that are later merged during the next meeting.
This evolution of the sketch can either happen on analog or digital media.
A tool supporting round-trip sketching should addresses these issues by providing means to capture context, manage revisions, and link sketches to related sketches or other artifacts.
It should then be possible to access this information using digital as well as augmented analog versions of a sketch.

Figure~\ref{fig:lifecycle} visualizes the conceptual workflow for round-trip sketching including different transitions from analog to digital media and back:
The lifecycle starts with the creation of an analog sketch (1).
The user decides that the sketch is worth capturing and adds a generated QR code label to the sketch (2) to be able to identify this version later.
Then, he or she uses a tablet to capture the sketch (3).
At that point, it is possible to add metadata like authors, textual annotations, or even short videos explaining the content of the sketch (4).
After the sketch is digitized, users may add content using tools like Gimp, Visio, or Photoshop, or utilize the sketch to redraw a revised digital version.
The tool should allow users to easily add these new revisions to the captured sketch.
Generally, if analog and digital versions of a sketch evolve concurrently, version control should help to keep track of their relation and should assist in merging versions existing in the analog as well as in the digital world.
Users may also link the whole sketch or parts of it to other related sketches or documents to embed the artifact in the work context (5).
To return from digital to analog, the user can print the revised sketch, for instance to bring it to a meeting (6).
Using the QR code on the printed or the initial sketch, the user can access all revisions of the sketch as well as linked artifacts and captured metadata (7).
To close the circle, users may add content to the printed sketch (8) and capture this new revision with a tablet to access, refine, or share it later (9). 

For a first prototype implementation supporting the workflow described above, we formulated six requirements:

\begin{figure*}[t]
\centering
    ~ %add desired spacing between images, e. g. ~, \quad, \qquad, \hfill etc. 
    \begin{subfigure}[b]{0.48\textwidth}
        \includegraphics[width=\textwidth]{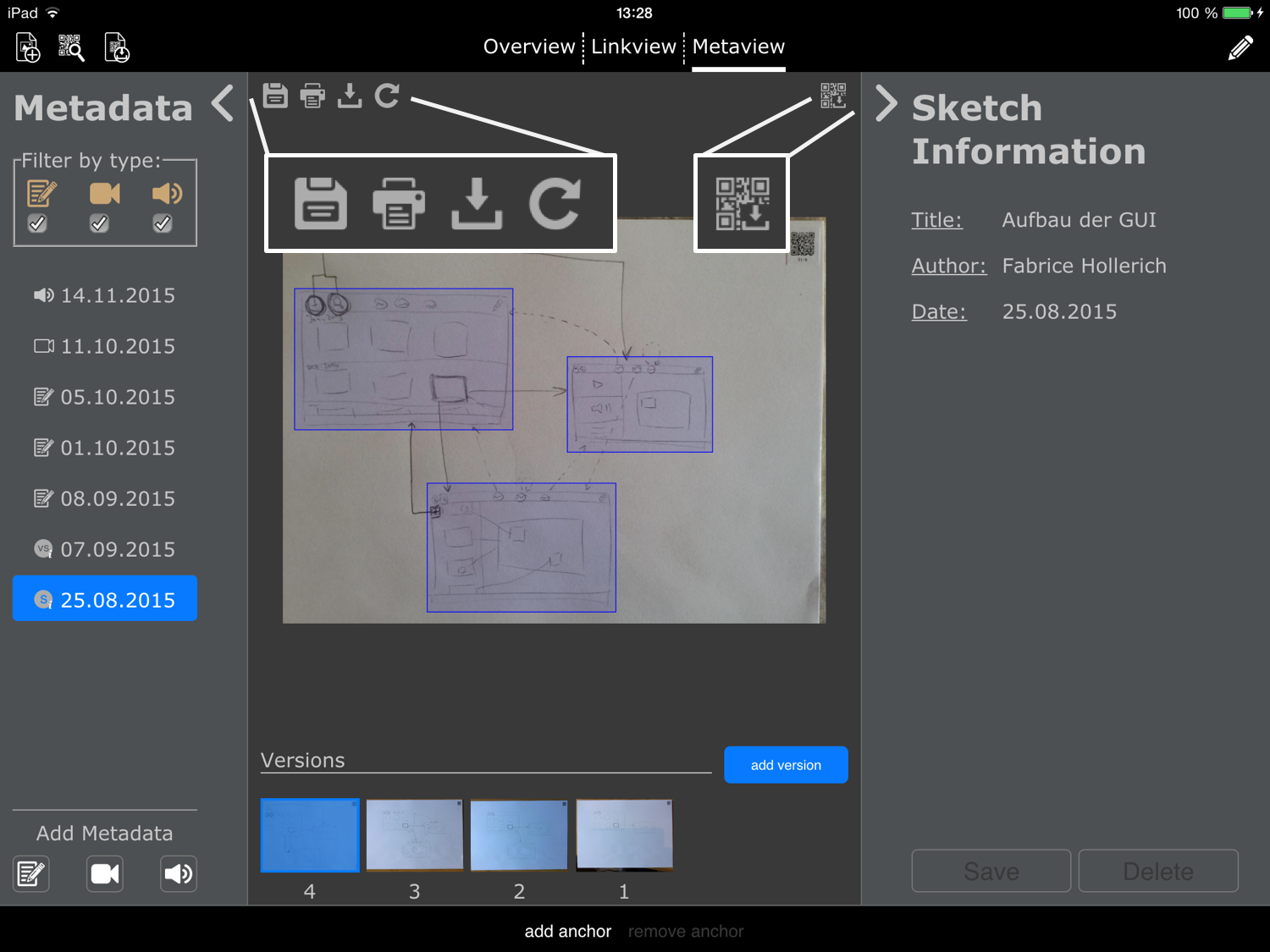}
        \caption{\emph{MetaView:} Show/add metadata, version control.}
        \label{fig:metaview}
    \end{subfigure}
    ~ %add desired spacing between images, e. g. ~, \quad, \qquad, \hfill etc. 
    \begin{subfigure}[b]{0.48\textwidth}
        \includegraphics[width=\textwidth]{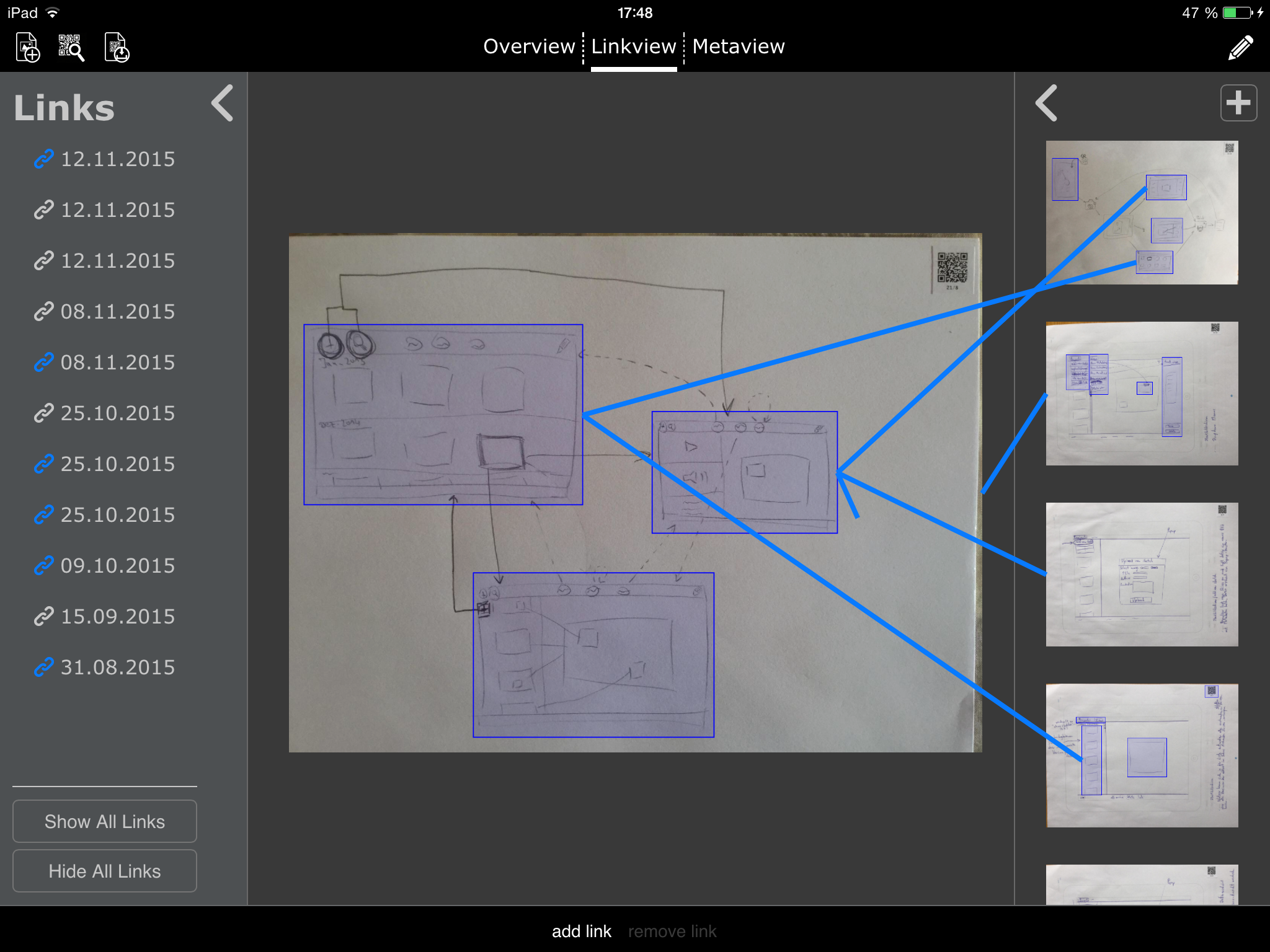}
        \caption{\emph{LinkView:}  Manage links to other sketches.}
        \label{fig:linkview}
    \end{subfigure}
    \caption{Two views of the \emph{LivelySketches} prototype.}
    \label{fig:views}
\end{figure*}

\vspace{-0.5\baselineskip}
% http://tex.stackexchange.com/a/29374/100364
\begin{itemize}[labelindent=0pt, labelwidth=\widthof{\emph{(augment)}}, label={\emph{(augment)}}, leftmargin=*, align=parleft, parsep=1ex, partopsep=1ex, topsep=1ex, itemsep=0.5ex]
\item[\textbf{REQ1:}\\\emph{(identify)}] The tool should allow users to unambiguously identify analog and digital sketches.
\item[\textbf{REQ2:}\\\emph{(capture)}] The tool should enable users to capture analog sketches along with context information. 
\item[\textbf{REQ3:}\\\emph{(version)}] The tool should enable users to add both analog and digital revisions to a captured sketch.
\item[\textbf{REQ4:}\\\emph{(link)}] The tool should allow users to link captured sketches to related sketches and other artifacts.
\item[\textbf{REQ5:}\\\emph{(print)}] The tool should allow users to print captured and digitally revised sketches.
\item[\textbf{REQ6:}\\\emph{(augment)}] The tool should allow showing captured metadata and linked artifacts of identified sketches.
\end{itemize}

These requirements cover the main requirements mentioned in the introduction, which we derived from related work and our own research on sketches and diagrams in software development (1$\rightarrow$REQ2+6, 2$\rightarrow$REQ3, 3$\rightarrow$REQ4, 4$\rightarrow$REQ1+5).

\begin{figure}[b]
\centering
\includegraphics[width=0.95\columnwidth]{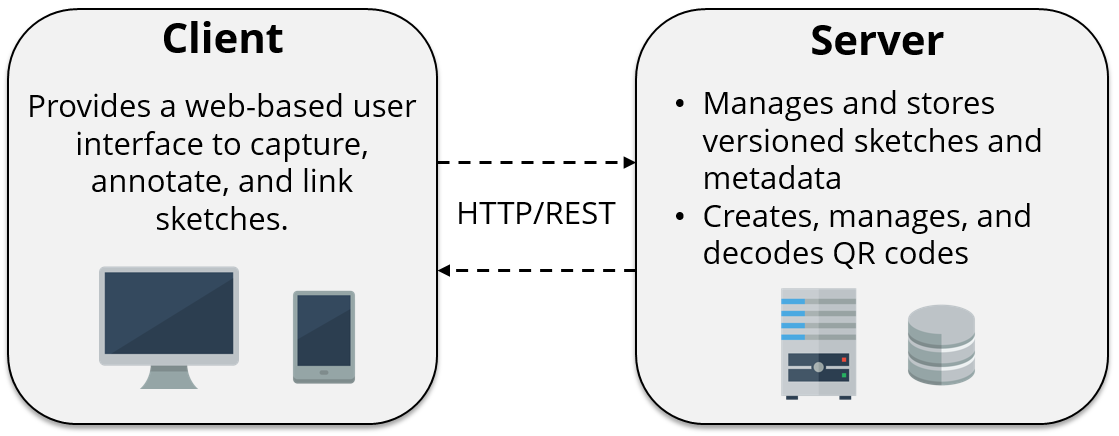}
\caption{Architecture of \emph{LivelySketches}}
\label{fig:architecture}
\end{figure}

%\begin{figure}[b]
%\centering
%\includegraphics[width=0.95\columnwidth]{figures/overview_icons.png}
%\caption{\emph{OverView:} Open or add sketch, scan QR code}
%\label{fig:overview}
%\end{figure}

\section{Prototype Implementation}
\label{sec:prototype-implementation}

To be able to evaluate the conceptual workflow, we created a prototype that implements the above requirements.
It uses a simple client-server architecture (see Figure~\ref{fig:architecture}).
The client is web-based and runs on both desktop and mobile browsers.
We optimized the GUI to be used on touch devices and tested it on an Apple iPad.
The server is responsible for storing and managing the versioned sketches and for creating and decoding the QR codes used to identify sketches.
It provides a REST API~\cite{Fielding00} that the client uses to upload and retrieve data.
The web client of \emph{LivelySketches} provides three main views: One view to open captured sketches and to add new ones (\emph{OverView}), one view to add and view metadata and to manage the revisions of a sketch (\emph{MetaView}), and one view to link a captured sketch or parts of it to other sketches (\emph{LinkView}).
The latter two views are depicted in Figure~\ref{fig:views}.
In the following, we describe how the prototype implements the above requirements.

\subsubsection*{REQ1 (Identification)}

%\begin{figure}[b]
%\centering
%\includegraphics[width=0.55\columnwidth]{qr-list}
%\caption{List of QR codes generated by \emph{LivelySketches}}
%\label{fig:qr-list}
%\end{figure}

\begin{figure}[t]
\centering
\includegraphics[width=\columnwidth]{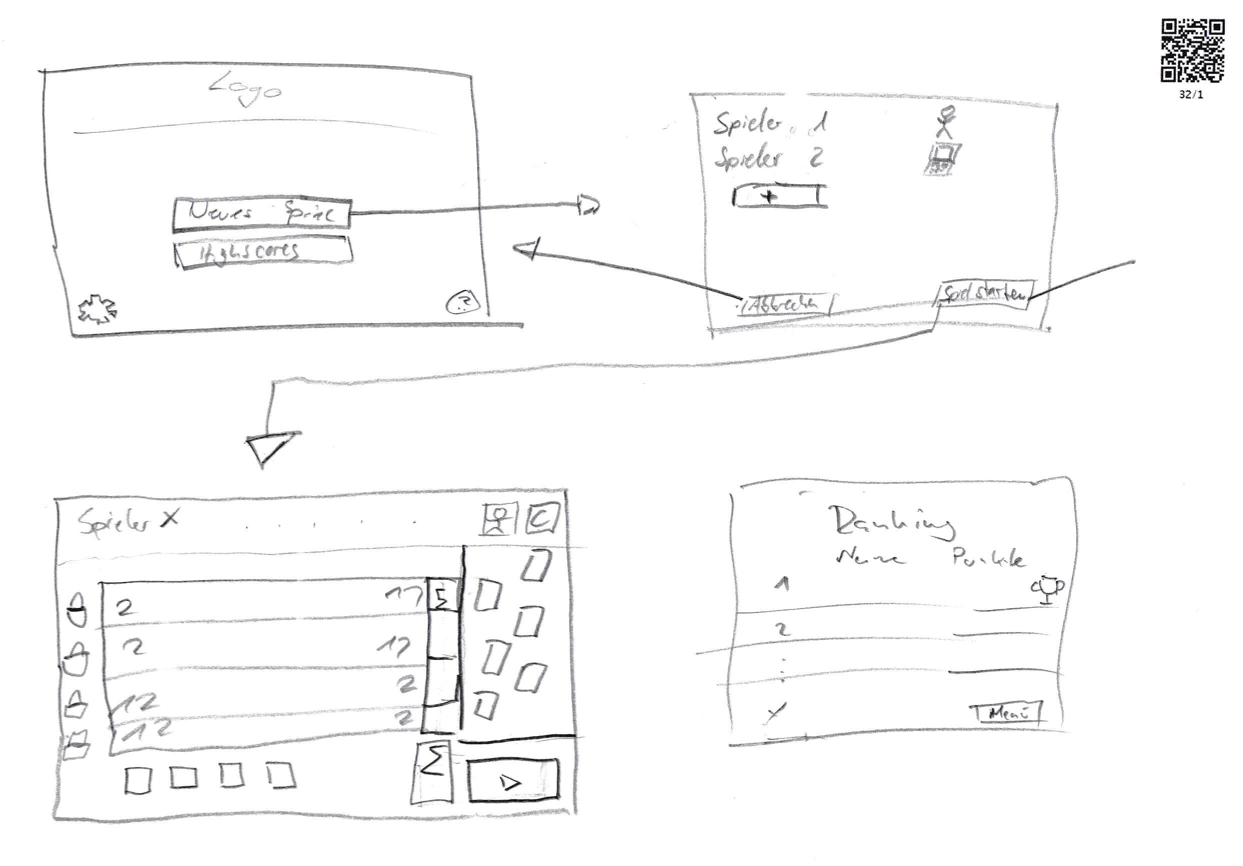}
\caption{Sketch with QR code created during formative study.}
\label{fig:study-sketch}
\end{figure}

Every sketch is identified by a \emph{Universally Unique Identifier} (UUID).
This identifier is either assigned to a sketch when it is uploaded to the server or by sticking a prepared QR code label to it.
\emph{LivelySketches} allows to create lists with QR codes that encode predefined UUIDs.
These lists can then be printed to adhesive labels.
The labels are very small ($1cm^2$) and thus do not distract from the main content of the sketch (see Figure~\ref{fig:study-sketch} for an example from the formative study).
One sheet contains 13 rows with different UUID and 10 identical labels in each row. % (see Figure~\ref{fig:qr-list} for an example).
This enables the user to mark different revisions of the same sketch drawn on different sheets of paper with the same label.
\emph{LivelySketches} then recognizes the identical codes and adds later uploaded sketches as revisions of the first one.
The tool also allows the user to generate labels for already captured sketches to mark new analog revisions of them.
Further, QR codes can be digitally added to a sketch after uploading it.
However, we recommend to always use labels to mark analog sketches to be able to reference them later.

\subsubsection*{REQ2 (Capturing)}

Using the main view of the app, the user can open already captured sketches or upload new sketches.
He or she can either upload a JPEG or PNG file or use the tablet's camera to take a picture of an analog sketch.
The tool then allows the user to add different meta information like title, author names, or date. 
After the sketch is uploaded, the \emph{MetaView} (see Figure~\ref{fig:metaview}) provides the functionality to add textual annotations, audio, or video files to the sketch.
Again, existing files can be uploaded or the tablet camera can be used to record these annotations.

\subsubsection*{REQ3 (Versioning)}

As mentioned above, sketches are often revised and redrawn.
To keep track of the history of a sketch, the prototype enables the user to add new revisions that were created either on analog or digital media.
\emph{LivelySketches} follows a state-based extensional versioning approach~\cite{Conradi98} that establishes a simple linear successor relationship between the revisions of a sketch.
The sequential order of the revisions is shown in the lower part of the \emph{MetaView}.
When adding a new revision, the tool asks for a commit message describing the modifications that took place between the two revisions.
Further, the tool assists the user in transferring  metadata from a previous version.

\subsubsection*{REQ4 (Linking)}

To link sketches, the prototype provides a dedicated view (see Figure~\ref{fig:linkview}).
A link always connects whole sketches or parts of sketches that are identified using a link anchor.
Currently, \emph{LivelySketches} only allows to use rectangular link anchors, but we plan to extend this with other shapes or a free-form selection to allow for a more fine-grained selection.

\subsubsection*{REQ5 (Printing)}

In the \emph{MetaView}, sketches can be either printed directly or the image files may be downloaded.
Furthermore, this view allows the user to print a list of QR code labels that may be used to identify new analog revisions of the sketch. 

\subsubsection*{REQ6 (Augmenting)}

Using the corresponding button in the header of \emph{LivelySketches}, it is possible to scan the QR code of an analog sketch.
The system then decodes the UUID of the sketch, opens the \emph{MetaView} and automatically switches to the most recent revision.
The user can then browse to the history of the scanned sketch, access or add metadata, or navigate to linked sketches.

\section{Formative User Study}

A formative user study is a study conducted ``during the development of a product [...] to mould or improve the product''~\cite{Travis06}.
To get early feedback for improving the prototype we conducted such a study with four participants.
All participants were computer science graduate students.
During the study, they worked in teams of two to design a graphical user interface for a dice game.
After a short introduction into the \emph{LivelySketches} prototype, we provided them the rules of the game they were going to design, paper and pencils for drawing sketches, an Apple iPad running the prototype, and a prepared sheet with QR code labels.
Then, we gave them a task description for the study, namely to design a GUI and a storyboard for the game at hand.
During the study, we captured audio and video data.
We recorded both the sketching activity on the desk with a camcorder as well as the interaction with the tool using a screen capturing software.
We provide all sketches created during the study as well as the screen captures as supplementary material~\cite{SupplementaryMaterial}.
After the study, the participants filled in a \emph{system usability scale} (SUS) questionnaire which we used as a ``quick and dirty'' way of accessing the tool's usability~\cite{Brooke96}.

Participants' comments and a SUS score of 81.3 indicate that they were very satisfied with the performance and stability of the tool.
Since we conducted the study to improve the tool, we will focus on possible improvements and feature requests in the following:
Because it is tedious to enter the author names each time a sketch is captured, participants requested a user management that would allow them to automatically add their own username or usernames of other registered authors.
Currently, all sketches are managed globally, which was not a problem during this short study, but for real-world usage, participants proposed to add a feature allowing to assign sketches to certain projects.

Regarding the link anchors, participants requested other shapes and free-form anchors to be more flexible in linking parts of sketches.
Further, they wanted to be able to automatically center the view around a selected anchor and then zoom into the linked part of the sketch.
This would be particularly useful when capturing large sketches, for example from whiteboards.
Regarding the linked sketches, participants proposed a global view visualizing all sketches and all links between them.
In this view, it would then be possible to zoom in and out and to navigate through this structure by following the links.
We will consider the comments and feature requests described above in the future development of the tool.

%\begin{figure}[t]
%\centering
%\includegraphics[width=0.85\columnwidth]{figures/study_setup_annotated}
%\caption{Setup of formative study}
%\label{fig:study-setup}
%\end{figure}

\section{Conclusion and Future Work}

In this paper, we presented the conceptual design of round-trip sketching as well as a prototype implementation named \emph{LivelySketches} that supports the lifecycle of sketches from analog to digital media and back.
It enables users to manually capture both analog and digital sketches as well as relevant context information.
The captured sketches can then be organized in a common version history.
Further, \emph{LivelySketches} allows linking sketches to related sketches.
We conducted a formative user study and got valuable insights to improve the tool.
In the future, we plan to extend \emph{LivelySketches} to allow users to also link sketches to other relevant resources like documents, emails, or source code~\cite{Baltes14a}.
A linked sketch could then be used to navigate through linked artifacts.

Although we tried to support a very common scenario motivated by related studies, the evolution of a sketch from paper to a digitally revised version (and back) is not the only possible workflow.
Walny et al.~\cite{Walny11a} present an overview of other possible lifecycles of software development sketches.
We want to evaluate the tool in a larger context to see how well it integrates into ``real-life'' settings.
The proposed approach works best with paper sketches, but it is also possible to capture sketches on whiteboards using a tablet camera.
In this scenario, the original sketch is lost when the whiteboard is erased, but the lifecycle can continue with the digitally captured version or with a printout.

%We focused mainly on software-related workflows, but 
\emph{LivelySketches} can be used to capture storyboards for graphical user interfaces, connect sketches and visualizations of different components of a software architecture, connect visualizations of dynamic and static aspects of software (e.g. UML class and sequence diagrams), or connect visualizations of different steps in the development process of a software project.
However, the application area of \emph{LivelySketches} is not limited to software development---the approach can be adapted to any discipline where sketching plays an important role.

\section*{Acknowledgments}

The authors would like to thank the participants of the formative study for their valuable feedback. % and Fabian Beck 



\bibliographystyle{IEEEtran}
\bibliography{IEEEabrv,literature}

\end{document}